\documentclass[11pt]{article}
\usepackage{amsmath,amssymb,epsf,color}

\makeatletter \@addtoreset{equation}{section} \makeatother

\def\ben{\begin{equation}}
\def\een{\end{equation}}

 \let\m=\mu \let\n=\nu  \let\p=\pi

\let\C=\Chi

 \def\bd{\begin{document}} \def\ed{\end{document}}
\def\ds{\documentstyle} \let\fr=\frac \let\bl=\bigl \let\br=\bigr
\let\Br=\Bigr \let\Bl=\Bigl
\let\bm=\bibitem
\let\na=\nabla
\let\pa=\partial \let\ov=\overline
\newcommand{\be}{\begin{equation}}
\newcommand{\ee}{\end{equation}}
\def\ba{\begin{array}}
\def\ea{\end{array}}
\def\ft#1#2{{\textstyle{\frac{\scriptstyle #1}{\scriptstyle #2} } }}
\def\fft#1#2{{\frac{#1}{#2}}}
\def\del{\partial}
\def\vp{\varphi}
\def\sst#1{{\scriptscriptstyle #1}}
\def\oneone{\rlap 1\mkern4mu{\rm l}}
\def\td{\tilde}
\def\wtd{\widetilde}
\def\ie{{\it i.e.\ }}
\def\dalemb#1#2{{\vbox{\hrule height .#2pt
        \hbox{\vrule width.#2pt height#1pt \kern#1pt
                \vrule width.#2pt}
        \hrule height.#2pt}}}
\def\square{\mathord{\dalemb{6.8}{7}\hbox{\hskip1pt}}}
\newcommand{\ho}[1]{$\, ^{#1}$}
\newcommand{\hoch}[1]{$\, ^{#1}$}
\newcommand{\bea}{\begin{eqnarray}}
\newcommand{\eea}{\end{eqnarray}}
\newcommand{\ra}{\rightarrow}
\newcommand{\lra}{\longrightarrow}
\newcommand{\Lra}{\Leftrightarrow}
\newcommand{\bp}{\tilde \beta^\prime}
\newcommand{\tr}{{\rm tr} }
\newcommand{\Tr}{{\rm Tr} }
\def\0{{\sst{(0)}}}
\def\1{{\sst{(1)}}}
\def\2{{\sst{(2)}}}
\def\3{{\sst{(3)}}}
\def\4{{\sst{(4)}}}
\def\5{{\sst{(5)}}}
\def\6{{\sst{(6)}}}
\def\7{{\sst{(7)}}}
\def\8{{\sst{(8)}}}
\def\n{{\sst{(n)}}}
\def\cA{{{\cal A}}}
\def\cB{{{\cal B}}}
\def\cF{{{\cal F}}}
\def\cH{{{\cal H}}}
\def\tV{\widetilde V}
\def\tW{\widetilde W}
\def\tH{\widetilde H}
\def\tE{\widetilde E}
\def\tF{\widetilde F}
\def\tA{\widetilde A}
\def\im{{{\rm i}}}
\def\tY{{{\wtd Y}}}
\def\ep{{\epsilon}}
\def\vep{{\varepsilon}}
\def\bD{{{\bar D}}}
\def\R{{{\mathbb R}}}
\def\C{{{\mathbb C}}}
\def\H{{{\mathbb H}}}
\def\CP{{{\mathbb C}{\mathbb P}}}
\def\RP{{{\mathbb R}{\mathbb P}}}
\def\Z{{{\mathbb Z}}}
\def\bA{{{\mathbb A}}}
\def\bB{{{\mathbb B}}}
\def\bC{{{\mathbb C}}}
\def\bD{{{\mathbb D}}}
\def\bE{{{\mathbb E}}}
\def\bZ{{{\mathbb Z}}}
\def\Re{{{\frak{Re}}}}
\def\Im{{{\frak{Im}}}}
\def\cosec{{\,\hbox{cosec}\,}}
\def\Gm{{\Gamma_{\!\! -}}}
\def\Gp{{\Gamma_{\!\! +}}}
\def\stan{{standard }}
\def\nonstan{{supernumerary }}
\def\p{{\partial}}
\def\kdel#1{{\fft{\del}{\del#1}}}

\def\bog{{Bogomolny }}

\def\beps{{\bar\epsilon}}
\newcommand{\ww}[1]{\\[0.#1cm]}
\def\eps{\epsilon}
\def\slashchar#1{\setbox0=\hbox{$#1$}           
   \dimen0=\wd0                                 
   \setbox1=\hbox{/} \dimen1=\wd1               
   \ifdim\dimen0>\dimen1                        
      \rlap{\hbox to \dimen0{\hfil/\hfil}}      
      #1                                        
   \else                                        
      \rlap{\hbox to \dimen1{\hfil$#1$\hfil}}   
      /                                         
   \fi}
\def\sd{\slashchar{D}}
\def\sp{\slashchar{\partial}}
\newcommand{\eq}[1]{(\ref{#1})}
\def\cdh{\Gamma\cdot H}
\def\cL{e^{-1} {\cal L}}

\begin{document}

\begin{flushright}
\end{flushright}

\vspace{25pt}
\begin{center}
{\large {\bf On Hybrid (Topologically) Massive Supergravity in Three
Dimensions}}

\vspace{15pt}

H. L\"u\hoch{1,2} and Yi Pang\hoch{3}

\vspace{10pt}

\hoch{1}{\it China Economics and Management Academy\\
Central University of Finance and Economics, Beijing 100081}

\vspace{10pt}

\hoch{2}{\it Institute for Advanced Study, Shenzhen University,
Nanhai Ave 3688, Shenzhen 518060}

\vspace{10pt}

\hoch{3} {\it Key Laboratory of Frontiers in Theoretical
Physics,\\Institute of Theoretical Physics, Chinese Academy of
Sciences, Beijing 100190, P.R.China}

\vspace{40pt}

\underline{ABSTRACT}
\end{center}

A class of hybrid (topologically) massive off-shell supergravities
coupled to an on-shell matter scalar multiplet was recently
constructed. The auxiliary field in the off-shell multiplet is
dynamical for generic values of the eight parameters. We find that
by choosing the parameters appropriately, it remains non-dynamical.
We perform linearized analysis around the supersymmetric AdS$_3$
vacuum and its Minkowski limit. The ghost-free condition for the
Minkowski vacuum is explored. For the AdS$_3$ vacuum, we obtain the
criticality condition and find that at the critical points, one of
the two massive gravitons becomes pure gauge and decouples from the
bulk physics, whilst the other has positive energy.  We demonstrate
that the mass of the BTZ black hole is non-negative at the critical
points. We also investigate general BPS solutions. For certain
parameter choices, we obtain exact solutions. In particular, we
present the BPS string (domain-wall) solution that is dual to
certain two-dimensional quantum field theory with an ultra-violet
conformal fixed point.

\vspace{15pt}

\thispagestyle{empty}




\newpage


\section{Introduction}

The foundation of Einstein's theory of gravity is the principle of
the general coordinate transformation invariance.  The
Einstein-Hilbert action is the minimal dynamical theory that
incorporates this principle. Its successes in large scale physics
notwithstanding, the theory is non-renormalizable in the frame work
of quantum mechanics.  Insisting on this principle, the
Einstein-Hilbert action can only be modified by adding
higher-derivative terms, such as polynomial invariants constructed
from the Riemann curvature tensor. While indeed theories with a
finite number of such higher-order terms can become renormalizable,
it is at the price of unitarity \cite{stelle}. At the level of
effective field theory, string provides an infinite number of
higher-order terms in such a specific way that the resulting theory
is expected to be both unitary and finite. However, this theory is
too complicated to play with.

Gravity in three dimensions is simple since the Einstein-Hilbert
action provides no physical degrees of freedom. However, by adding
higher derivative terms dynamics can be generated. The resulting
theory becomes non-trivial but simpler than Einstein gravity in four
dimensions, thus provides a non-trivial toy model for studying
quantum gravity. The best known example is topologically massive
gravity, which is constructed by adding a Lorentz-Chern-Simons term
to the Einstein-Hilbert action. Consequently, a massive graviton
emerges \cite{Deser:1981wh,tmg}. The theory is not unitary with the
standard Einstein-Hilbert action, but can be made so by reversing
the sign of the action. However, when coupled to a cosmological
constant \cite{deser2}, the negative sign of the Einstein-Hilbert
action implies that the BTZ black hole, which is an excited state in
the theory, has negative mass. In \cite{lisost}, cosmological
topologically massive gravity with the standard Einstein-Hilbert
action was revisited. It was argued that the ghost-like massive
graviton decouples at a certain critical point of the parameter space.
The critical theory is then conjectured to be self-consistent
quantum gravity {\it via} the AdS/CFT correspondence.

Subsequently, new massive three-dimensional gravity (NMG) with
quadratic Riemann curvature invariants was constructed and shown to
be unitary \cite{bht1}. This inspires later constructions of more general
(topologically) massive (super)gravities \cite{Andringa:2009yc,Bergshoeff:2010mf} in three dimensions, as
well as the higher dimensional generalizations \cite{Lu:2010sj}.
(See also a recent review \cite{Bergshoeff:2010iy}.) All these
theories involve only the metric (or the supergravity multipet), and
the most general construction has seven parameters
\cite{Bergshoeff:2010mf}. Recently, an eight-parameter ${\cal N}=1$
supergravity with a matter scalar multiplet was constructed
\cite{lps}.  The theory is of particular interest since it is hybrid
in that the supergravity multiplet is off-shell with higher
derivative terms in the action whilst the matter multiplet is
on-shell with at most two derivatives.

   An important feature of all these three-dimensional massive
supergravities is that the supersymmetry can be off-shell. The
consequence is that such a theory can be complete in terms of
supersymmetry by adding just any finite number of higher-derivative
super invariants. This is very different from string theory or
M-theory whose supersymmetry is realised on-shell: the completeness of the supersymmetry alone requires an infinite number of higher-order
terms once just one term beyond the second derivative is introduced.
(See for example \cite{Bergshoeff:1989de}.) Furthermore, the
existence of the hybrid theory \cite{lps} demonstrates that in three
dimensions, even if higher-derivative terms in gravity sector are
inevitable in a quantum theory, the matter sector can still be
minimally coupled (In stringy frame). This is desirable since the matter sector such as the Standard Model is indeed renormalizable without having to go
beyond two derivatives. These features make three dimensions an
attractive starting point to study quantum gravity.

   We begin our discussion in section 2 by reviewing hybrid
topologically massive supergravity.  We give the bosonic action, the
equations of motion and the supersymmetry transformation rules. As
in typical off-shell supergravity, the auxiliary field can acquire
dynamics when higher-order super invariants are added to the action.
This makes the analysis much more complicated. In section 3, we find
that by choosing the eight parameters appropriately, the auxiliary
scalar field $S$ stays non-dynamical. This simplifies the theory
significantly, but it is non-trivial with five parameters left.
Interestingly, it turns out that the ratio of the coefficients of
the $R^2$ and $R^{\mu\nu}R_{\mu\nu}$ terms is $-3/8$, the number
needed for the new massive gravity \cite{bht1} to be unitary.  We
also present an analogous four-parameter pure massive supergravity
without the matter multiplet.

For theories with higher derivative terms,  it is typical that
massive modes with with negative energy can arise.  Since the matter multiplet does not involve higher derivative terms, the problem of having negative energy modes lies mainly in the gravity sector.  In particular the analysis for the traceless spin-2 graviton mode is identical to that for the pure massive supergravity.  Thus our analysis and many conclusions applies for the four-parameter pure massive supergravity.

We first consider the limit of the five-parameter hybrid theory for which
its supersymmetric vacuum becomes Minkowski space-time, and then
perform a linearized analysis around this vacuum. The spectrum
consists of one massless scalar mode and two massive graviton modes.
The consequence of the inclusion of the scalar multiplet is that the
theory in general becomes non-unitary regardless of the sign of the
Einstein-Hilbert action. However, we find that there exists a
special point in parameter space for which the on-shell Hamiltonian
for the two massive graviton modes vanishes identically, signalling
that the linearized analysis breaks down, and the possibility that the
theory may become unitary.

We also perform a linearized analysis around the supersymmetric
AdS$_3$ vacuum.  Some aspects of the linear analysis for the
tensorial and scalar modes for general parameters were given in
\cite{lps}.  Specializing to our choice of parameters, we find that
the theory in general has one scalar mode and two massive graviton
modes.  Since the matter scalar mode cannot be gauged away, we have
to restrict parameters so that the scalar mode is not ghost-like.
This can be achieved by requiring that the super invariant in the
action involving the Einstein-Hilbert term has to have positive
coefficient.  (This provides an extra constraint than the pure gravity theory.) Then one of the massive gravitons becomes inevitably
ghost-like. We find the critical points for which the ghost massive
graviton becomes pure gauge and decouples from the bulk physics and
then verify that the remaining massive graviton indeed has positive
energy. This result applies also to the four-parameter pure massive
supergravity.  Owing to the complexity arising from the dynamical
nature of the auxiliary field $S$ for generic parameters, no
conclusion was made in \cite{lps} about the stability of the scalar
mode. For our specialized theory where $S$ is non-dynamical, we
demonstrate that the scalar mode is stable at the critical points
satisfying the Breitlohner-Freedman bound. This suggests that the
theory may be well-defined at these critical points.

    In section 4, we obtain the mass and angular momentum
for the BTZ black hole that is asymptotic to the supersymmetric
AdS$_3$ vacuum.  We demonstrate that at the critical points, the
mass is non-negative and always greater or equal to the angular
momentum.  We also verify that the first law of thermodynamics
holds.

   It should be emphasized that although our analysis focused
on the hybrid theory, the linearized analysis of the traceless graviton mode and the BTZ energy calculation apply equally well to the four-parameter pure massive gravity.  The critical points for the general massive pure supergravity were obtained in \cite{Bergshoeff:2010mf}. Our examination of the energy of the remaining non-trivial massive graviton and the mass of the BTZ black hole reveals additional properties of the theory at the critical points.  The inclusion of the on-shell matter multiplet does not alter these results, but instead provides further constraints on the parameter space.

    In section 5, we study BPS string (domain-wall) solutions
of the eight-parameter supergravity. The equations reduce to one
third-order ordinary non-linear differential equation.  For certain
parameter choices, including the five-parameter theory, the
solutions can be obtained explicitly.  These solutions are
asymptotic to AdS$_3$ and are expected to be dual to certain
two-dimensional field theory with an ultra-violet conformal fixed
point. We discuss the characteristics of the spectrum using the
standard free-scalar approach.

  In section 6, we investigate general BPS solutions.  We find
that for general choice of the parameters, equations can be reduced to two differential equations. For some special choices of parameters, we
are able to obtain the exact solutions.  In particular we derive
solutions that arise at the critical points of the five-parameter
theory discussed in section 3. The paper concludes in section 7.

\section{The theory}

Hybrid ${\cal N}=1$ topologically massive supergravity involves an
off-shell supergravity multiplet $(e^a_\mu,\psi_\mu, S)$ and an
on-shell scalar matter multiplet $(\phi,\psi)$.  The full action up
to quartic fermion were given in \cite{lps}. For our purpose, we are
concerned with the bosonic sector, namely
\begin{eqnarray}
I &=& \fft{1}{2\kappa^2} \int d^3 x \sqrt{-g}\Big[\sigma
e^{-2\phi}[R+4(\partial\phi)^2+4m^2+2S^2]+ 4\tilde m S -2a (R S +
2S^3)\cr
&& + \ft14\alpha(4R_{\mu\nu}R^{\mu\nu}-R^2-8 (\partial S)^2+
12S^4+4RS^2)+ c(3RS^2+10S^4)\cr
&&+b(R^2-16(\partial S)^2+12RS^2+36S^4)\Big] + 2\beta m {\cal
L}_{\rm LCS}\,,\label{genaction}
\end{eqnarray}
where
\begin{equation}
{\cal L}_{\rm LCS}=\ft12 \epsilon^{\lambda \mu\nu}
\Gamma^{\rho}_{\lambda\sigma} \Big(\partial_\mu
\Gamma^{\sigma}{}_{\rho\nu} + \ft{2}{3} \Gamma^{\sigma}{}_{\mu\tau}
\Gamma^{\tau}{}_{\nu\rho}\Big)\,.
\end{equation}
The theory contains eight parameters, including seven continuous
ones $(m,\tilde m, \alpha, \beta, a, b, c)$ and one discrete
$\sigma$ which takes values of $0, \pm 1$.  We do not count the
three dimensional $\kappa$ as a non-trivial parameter. The
supersymmetric transformation rules are given by \cite{lps}
\begin{eqnarray}
&&\delta e^a_\mu = \ft12 \bar \epsilon \gamma^a \psi_\mu\,,\qquad
\delta \psi_\mu = D_\mu \epsilon+ \ft12 \gamma_\mu S
\epsilon\,,\qquad \delta S= \ft14 \bar\epsilon \gamma^{\mu\nu}
D_\mu\psi_\nu - \ft14 \bar \epsilon \gamma^\mu \psi_\mu S\,;\cr 
&&\delta \psi = \ft14 e^{\fft54\phi}(\gamma^\mu \partial_\mu \phi+ S
+ m) \epsilon\,,\qquad \delta \phi = e^{-\ft54\phi} \bar \epsilon
\psi\,. \label{susytrans}
\end{eqnarray}
The first three transformations are self contained and close
off-shell; the last two transformations close on-shell.  As
discussed in \cite{lps}, one cannot truncate out the scalar
multiplet to obtain the seven-parameter theory.  The supersymmetry
transformation rules imply that setting $(\phi,\psi)$ zero, the
auxiliary field $S$ is fixed and the whole theory reduces to
standard cosmological topologically massive supergravity\cite{Deser:1982sw}.

The equation of motion associated with variation of the dilaton is
given by
\begin{equation}\label{phieom}
4\Box\phi - 4 (\del\phi)^2 + R + 2 S^2 + 4 m^2 =0\,.
\end{equation}
The equation of motion for the auxiliary field $S$ is given by
\begin{equation}\label{seom}
(\alpha + 8b)\Box S + \sigma e^{-2\phi} S + \tilde m -\ft12 a (R + 6
S^2) + (3\alpha + 36 b + 10 c) S^3 + \ft12 (\alpha + 12b + 3c)
RS=0\,.
\end{equation}
The Einstein's equations are more complicated, given by
\begin{eqnarray}
&&\sigma e^{-2\phi}(R_{\mu\nu} + 2 \nabla_\mu\nabla_\nu \phi) -
2\tilde m S
g_{\mu\nu}\cr 
&&+\alpha\Big[\Box R_{\mu\nu} - \ft12 \nabla_\mu \nabla_\nu R - 4
R_{\mu}{}^\lambda R_{\lambda\nu} + \ft52 R R_{\mu\nu} + \ft32 g_{\mu
\nu} (R_{\rho\sigma}R^{\rho\sigma} - \ft7{12} R^2)\cr 
&&\qquad -\ft32S^4 g_{\mu\nu} + G_{\mu\nu} S^2 - (\nabla_\mu
\nabla_\nu - g_{\mu\nu}\Box)S^2 - 2\partial_\mu S\partial_\nu S +
(\partial S)^2 g_{\mu\nu}\Big]\cr 
&&-2\beta m C_{\mu\nu} + 2 a \Big[S^3 g_{\mu\nu} - G_{\mu\nu} S +
(\nabla_\mu\nabla_\nu -
g_{\mu\nu}\Box) S\Big]\cr 
&&-b\Big[(\nabla_\mu \nabla_\nu -g_{\mu\nu}\Box)F - F R_{\mu\nu}+
16\partial_\mu S\partial_\nu S +\ft12 g_{\mu\nu}(R^2 - 16 (\partial
S)^2 + 12 R S^2 + 36 S^4)\Big]\cr 
&&+ c\Big[3S^2 R_{\mu\nu} -3(\nabla_\mu\nabla_\nu - g_{\mu\nu}\Box)
S^2 - \ft12g_{\mu\nu}(3RS^2 + 10 S^4)\Big]=0\,, \label{einstein}
\end{eqnarray}
where
\begin{eqnarray}
G_{\mu\nu} &=& R_{\mu\nu} - \ft12 R g_{\m\nu}\,,\cr 
C_{\mu\nu} &=& \epsilon_\mu{}^{\rho\sigma}\nabla_{\rho}
(R_{\sigma\nu} -\ft14 g_{\sigma\nu} R)\,,\cr 
F&=&2(R + 6 S^2)\,.
\end{eqnarray}
Note that the supergravity multiplet involves up to four derivatives
whilst the matter scalar $\phi$ involves at most two derivatives.

\section{Super-NMG with scalar multiplet and CS terms}

As we can see in the previous section, the auxiliary field $S$ in
the generalized topologically massive gravity acquires dynamical
terms when higher-order off-shell super invariants are involved.
However, for some specific choice of parameters, we find that the
$S$ field remains non-dynamical.  This corresponds to set
\begin{equation}
\alpha = 6c\,,\qquad a=0\,,\qquad b=-\ft34 c\,.\label{auxcon}
\end{equation}
The reduced theory has five parameters, and the action is given by
\begin{eqnarray}
I &=& \fft{1}{2\kappa^2}\int d^3x \sqrt{-g}\Big[\sigma e^{-2\phi} (R
+ 4 (\del\phi)^2 + 4m^2 + 2S^2) + 4\tilde m S + \fft{1}{6\nu^2} S^4
\cr 
&&+ \fft{1}{\nu^2} (R_{\mu\nu}R^{\mu\nu} - \ft38 R^2)\Big] +
\fft{1}{\mu} {\cal L}_{\rm LCS}\,.\label{newlag}
\end{eqnarray}
Note that here we have renamed two parameters, namely $\nu^2=1/(6c)$
and $\mu=1/(2\beta m)$. The parameters $(m,\tilde m,\mu,\nu)$ all
have the same dimension $[\mbox{length}]^{-1}$. The parameter $\nu$
is chosen for the convenience of dimensional analysis and it is
understood that $\nu^2$ can be negative as well.

The equation of motion for $S$ now becomes purely algebraic, namely
\begin{equation}
\sigma S e^{-2\phi} + \tilde m + \fft{1}{6\nu^2} S^3=0\,.
\end{equation}
As in the general case, the supersymmetric vacuum is an AdS$_3$
\cite{lps}, but now with
\begin{equation}
R_{\mu\nu}=-2m^2 g_{\mu\nu}\,,\qquad S=-m\,,\qquad \phi=0\,,\qquad
\tilde m=-\sigma m - \fft{m^3}{6\nu^2}\,.\label{ads3}
\end{equation}
Note that in this paper, we shall take a convention that $\phi=0$
for the AdS$_3$ vacuum.  This can always be achieved if we let the
parameter $\sigma$ to be continuous.  When $m=0=\tilde m$, the
vacuum becomes Minkowski space-time.

   It is interesting to note that the $-\ft38$ factor of ghost-free new
massive gravity \cite{bht1} also arises in our case. Thus the theory
is a hybrid generalization of the pure super-NMG.  It is worth
pointing out that all the off-shell super invariants in
(\ref{genaction}) that decouple from the dilaton $\phi$ are the
exactly the same as those in pure massive supergravity constructed
in \cite{Bergshoeff:2010mf}. This implies that the specialization we
obtain in our hybrid theory also exists in the pure supergravity
theory.  It is given by
\begin{eqnarray}
I &=& \fft{1}{2\kappa^2}\int d^3x \sqrt{-g}\Big[\sigma (R - 2S^2) +
4\tilde m S + \fft{1}{6\nu^2} S^4 \cr 
&&+ \fft{1}{\nu^2} (R_{\mu\nu}R^{\mu\nu} - \ft38 R^2)\Big] +
\fft{1}{\mu} {\cal L}_{\rm LCS}\,.\label{newlag1}
\end{eqnarray}
The theory has four non-trivial parameters. The fermionic action can
be read off from \cite{Bergshoeff:2010mf}. The off-shell
supersymmetric transformation rule is given by
\begin{equation}
\delta e^a_\mu = \ft12 \bar \epsilon \gamma^a \psi_\mu\,,\qquad
\delta \psi_\mu = D_\mu \epsilon+ \ft12 \gamma_\mu S
\epsilon\,,\qquad \delta S= \ft14 \bar\epsilon \gamma^{\mu\nu}
D_\mu\psi_\nu - \ft14 \bar \epsilon \gamma^\mu \psi_\mu S\,.
\end{equation}
It should be emphasized that this pure super-NMG cannot be
obtained by truncating out the the scalar multiplet from
(\ref{newlag}).

Although our primary focus is on the hybrid theory (\ref{newlag}),
many of the results can also apply to the pure gravity theory
(\ref{newlag1}). This is because the scalar matter mulitplet involves only two derivatives and hence there is no massive mode with negative energy associated with the scalar multiplet.  For the tensorial modes, as we shall discuss later, with appropriate gauge choice, the linear analysis is exact the same for both theories.

\subsection{The Minkowski limit}

Let us first set $m=0=\tilde m$ so that the supersymmetric vacuum is
three-dimensional Minkowski space-time.  We consider linearized
excitations around the vacuum
\begin{equation}
g_{\mu\nu}=\eta_{\mu\nu} + h_{\mu\nu}\,,\qquad S=\bar S+ s\,, \qquad
\phi\rightarrow \bar \phi + \phi\,,
\end{equation}
where $\bar S=0=\bar \phi$. We may adopt the following gauge for the
metric \cite{ Andringa:2009yc,sdesergauge}
\begin{equation}
h_{ij}=-\varepsilon^{ik}\varepsilon^{jl} \frac{\partial_{k}
\partial_{l}}{\nabla^2}\varphi\,,\qquad
h_{0i}=-\varepsilon^{ij}\frac{1}{\nabla^2}\partial_{j}\xi\,,\qquad
h_{00}=\frac{1}{\nabla^2}(N+\Box\varphi)\,.
\end{equation}
This gauge choice amounts to
\begin{equation}
\partial_ih_{ij}=0\,,\qquad\partial_ih_{0i}=0\,.
\end{equation}
It should be emphasized that in this special gauge, $(\varphi, \xi,
N)$ are exactly the three gauge-invariant quantities constructed
from the metric.

   The quadratic action for perturbations around the vacuum is given
by
\begin{eqnarray}\label{qlag1}
I^{(2)}&=&\int d^3x\Big\{
-\sigma\Big(\ft12(N\varphi+\varphi\Box\varphi-\xi^2)
+4\phi\Box\phi+2\phi N+4\phi\Box\varphi\Big)\cr 
&& -\frac{1}{2\mu}N\xi+\frac{1}{2\nu^2}\xi
\Box\xi+\frac{1}{8\nu^2}N^2\Big\}\,.
\end{eqnarray}
The equation of motion for $N$ is purely algebraic and can be solved
explicitly, given by
\begin{equation}
N=4\nu^2 (\fft{1}{2\mu} \xi + 2 \sigma \phi + \ft12 \sigma
\varphi)\,.
\end{equation}
Sustituting this into (\ref{qlag1}), we have
\begin{eqnarray}
I^{(2)} &=& \int d^3x (T - V)\,,\cr 
T&=& -\sigma (\ft12 \varphi\Box\varphi +4\phi\Box\phi+
4\phi\Box\varphi) + \frac{1}{2\nu^2}\xi\Box\xi\,,\cr 
V&=& -\ft12\sigma\xi^2 +2\nu^2(\frac{1}{2\mu}\xi +2\sigma\phi+\ft12
\sigma \varphi)^2\,.
\end{eqnarray}
It is clear that if we diagonalize the kinetic terms of
$(\phi,\varphi)$, there is a ghost field regardless the sign of the
parameter $\sigma$. Since the kinetic terms always involve both
ghost and non-ghost fields, we cannot diagonalize the kinetic and
mass terms simultaneously. Nevertheless we can analyze its spectrum
by examining the equations of motion.  They are given by
\begin{eqnarray}
2\Box\phi + \Box\varphi + 2\nu^2 \Big(\fft{1}{2\nu} \xi + 2\sigma
\phi + \ft12\sigma \varphi\Big) &=& 0\,,\cr 
4\Box\phi + \Box \varphi + 2\nu^2\Big(\fft{1}{2\nu} \xi + 2\sigma
\phi + \ft12\sigma \varphi\Big) &=& 0\,,\cr 
\fft{1}{\nu^2} \Box\xi + (\sigma -\fft{\nu^2}{\mu^2})\xi -
\fft{2\nu^2\sigma}{\mu} (2 \phi + \ft12 \varphi) &=& 0\,.
\end{eqnarray}
In the momentum space
\begin{equation}
    \varphi=\int d^3\vec{p}(\varphi_{p}e^{-ipx}+c.c)\,,\quad \phi=\int
    d^3\vec{p}(\phi_{p}e^{-ipx}+c.c)\,,\quad\xi=\int
    d^3\vec{p}(\xi_{p}e^{-ipx}+c.c)\,,
\end{equation}
the equations of motion become
\begin{eqnarray}
M^2(2\phi_p+\varphi_p)+2\nu^2(\frac{1}{2\mu}\xi_p+
2\sigma\phi_p+\ft12\sigma\varphi_p)&=&0\,,\cr
M^2(\varphi_p+4\phi_p)+2\nu^2(\frac{1}{2\mu}\xi_p
+2\sigma\phi_p+\ft12\sigma\varphi_p)&=&0\,,\cr 
(\frac{M^2}{\nu^2}+\sigma-\frac{\nu^2}{\mu^2})
\xi_p-\frac{2\nu^2\sigma}{\mu}(2\phi_p+\ft12\varphi_p) &=&0\,.
\end{eqnarray}
where $M^2\equiv (p^{0})^2-\vec{p}\cdot\vec p$ is the mass-squared
parameter.  There are three non-trivial solutions.  The first has
vanishing $M$ with
\begin{equation}
\varphi_{p}=-4\phi_{p}\,,\qquad \xi_{p}=0\,.
\end{equation}
The corresponding on-shell Hamiltonian is given by
\begin{equation}
    H=4\sigma\int d^3x(\dot{\phi}^2+(\nabla\phi)^2).
\end{equation}
Thus $\sigma=-1$ gives rise to a massless ghost scalar.  The absence
of such a ghost requires that $\sigma\ge 0$.

     The other two solutions are given by
\begin{equation}
M^2_{\pm}=\frac{\nu^2}{2}\Big[\frac{\nu^2}{\mu^2}-2\sigma\pm
\sqrt{\frac{\nu^2}{\mu^2}(\frac{\nu^2}{\mu^2}-4\sigma)}\Big]\,,
\qquad \phi_p=0\,,\qquad \xi_p=-\frac{\mu}{\nu^2}
(M^2_{\pm}+\sigma\nu^2)\varphi_p\,.
\end{equation}
The absence of tachyon modes is guaranteed by
\begin{equation}\sigma>0\,,\quad\nu^2\in(-\infty,0)\cup
(4\sigma\mu^2,+\infty)\qquad
\hbox{or}\qquad\sigma<0\,,\quad\nu^2\in(-\infty,4\sigma\mu^2)
\cup(0,+\infty)\,. \label{munucon}
\end{equation}
In the limit $\nu\rightarrow\infty$, $M_{+}$ becomes infinity and
the corresponding mode decouples. In addition, $M_{-} \rightarrow
(\sigma\mu)^2$ and hence the corresponding mode is exactly the
massive graviton discussed in \cite{Deser:1981wh}. At the first
sight, the massive modes are scalars, but a careful study of the
supersymmetry transformation rules shows that they are in fact
spin-2 particles \cite{Andringa:2009yc,Bergshoeff:2010mf}. Thus the
spectrum consists of one massless scalar modes and two massive
graviton modes.

The on-shell Hamiltonian for the massive gravitons are given by
\begin{eqnarray}
H_{\pm}&=&\ft12\Big(\frac{\mu^2}{\nu^6}(M^2_{\pm}+
\sigma\nu^2)^2-\sigma\Big) (\dot{\varphi}^2+(\nabla\varphi)^2)\cr
&& + \Big(\fft{M_\pm^4}{2\nu^2} - \fft{\sigma \mu^2}{2\nu^4}
(M^2_\pm + \sigma \nu^2)^2\Big)\varphi^2\,.
\end{eqnarray}
It is easy to see that for $\sigma <0$, both kinetic terms are
non-negative. The theory has one ghost-like massless scalar and two
ghost-free massive gravitons.  For $\sigma>0$, the kinetic term in
$H_-$ is negative, and the theory has a well-defined massless
scalar, but one of the two gravitons is ghost like. Note that for
pure new massive supergravity with no matter scalar multiplet, the
theory is ghost free when $\sigma <0$.

     When the condition (\ref{munucon}) is saturated, namely
\begin{equation}
\nu^2=4\sigma \mu^2\,,
\end{equation}
both Hamiltonian $H_\pm$ vanish, signaling that the linearized
analysis breaks down and suggesting a possibility that the theory
might become ghost free. Of course, to determine this definitively,
higher-order interactions become non-negligible and a
non-perturbative analysis may be necessary.  We shall not proceed in
this direction here.

\subsection{Linearization around the AdS$_3$}

      The linear analysis in Minkowski space-time demosntrates that
there is at least one ghost-like field in the spectrum.  It is then
of interest to study the linear perturbation in the supersymmetric
AdS$_3$ to investigate whether there exists a critical point where
this ghost field decouples from the bulk physics, as in chiral
gravity proposed in \cite{lisost}.  Since the matter scalar $\phi$
has the standard dynamics, there can be no critical points
associated with the scalar and its physical degree of freedom cannot
be gauged way. It is thus necessary to require that $\sigma>0$.
 As we see in the flat space-time analysis, the
choice of $\sigma>0$ implies that one of the two massive gravitons
is ghost like.  We are interested in finding critical points to
gauge away this ghost graviton, but still to keep the other one so
that bulk gravity is non-trivial.

    The linear perturbation of the eight-parameter theory
(\ref{genaction}) was analyzed in \cite{lps}.  Owing to the
complexity of the theory and the dynamical nature of the auxiliary
field $S$, there was not a concrete conclusion for the scalar
perturbation.  The situation becomes much simpler for the reduced
action (\ref{newlag}) for which the auxiliary field $S$ is
non-dynamical.

As in \cite{lps}, we expand the metric around the supersymmetric
AdS$_3$ background (\ref{ads3}) as $g_{\mu\nu} = \bar g_{\mu\nu} +
h_{\mu\nu}$ and impose the gauge condition
\begin{equation}
\nabla^{\mu} (h_{\mu\nu} - \ft13 \bar g_{\mu\nu} h)=0\,,\qquad
\hbox{where}\qquad h\equiv \bar g^{\mu\nu} h_{\mu\nu}\,.
\end{equation}
The scalar fields $(S,\phi)$ are expanded around the symmetric
solution $\bar S=-m$ and $\bar \phi=0$, and we denote the
fluctuation fields by $s$ and $\phi$, respectively.  Setting $m=1$
for simplicity, the equations for the scalar modes are given by
\begin{eqnarray}
6\Box\phi-(\Box-3)h-6s&=&0\,, \cr 
(\sigma+\fft{1}{2\nu^2})s+2\sigma\phi &=& 0\,,\cr 
(\fft{1}{2\nu^2}-\sigma)(\Box-3)h+
12\sigma(\Box-3)\phi&=&0\,.\label{scalar1}
\end{eqnarray}
Note that the parameter $\mu$ does not enter the equations of the
scalar modes.  Depending on the relation between the parameters
$\sigma$ and $\nu$, four classes of solutions may emerge.

Firstly, when $\sigma = - 1/(2\nu^2)$, the equations in
(\ref{scalar1}) can be reduced to
\begin{equation}
   (\Box-3)h=0\,,\qquad \phi=0\,,\qquad s=0.
\end{equation}
This is exactly the same as the scalar perturbation of the pure
AdS$_3$ gravity, and hence it can be easily shown to be a pure
gauge.  Secondly, for $\sigma=1/(2\nu^2)$, (\ref{scalar1}) reduces
to
\begin{equation}
(\Box-3)h - 24\phi=0\,,\qquad (\Box-3)\phi=0\,.
\end{equation}
The third class corresponds to $(\Box-3)h=0$ and $(\Box-3)\phi=0$.
This requires that $\sigma = -3/(10\nu^2)$ and $s=3\phi$. Although
$h$ is pure gauge in this case, the free scalar $\phi$ is
non-trivial.   The remaining fourth class corresponds to generic
values of $\sigma$ and $\mu$.  Since we are looking for solutions
where $h$ is non-trivial, the last equation in (\ref{scalar1})
implies that
\begin{equation}
h=\fft{24\sigma \nu^2}{2\sigma \nu^2 -1} \phi\,.
\end{equation}
We then find
\begin{equation}
\Big(\Box - \fft{8\sigma \nu^2(1+4\sigma\nu^2)}{(1+2\sigma\nu^2)^2 }
\Big)\phi=0\,,\qquad s=\fft{-4\sigma\nu^2}{2\sigma\nu^2+1}\phi\,.
\end{equation}
Tne Breitlonhner-Freedman bound in 3-dimensions gives
\begin{equation}
\fft{8\sigma
\nu^2(1+4\sigma\nu^2)}{(1+2\sigma\nu^2)^2}\geq-1\qquad\Longrightarrow
\qquad\nu^2(\sigma+3\sigma^2\nu^2)\geq-\frac{1}{12}\,.\label{bfcon}
\end{equation}

The equations for the tensor modes are determined by the tracelss
part of the Einstein equations.  It was shown in \cite{lps} that
these equations are the same as the seven-parameter pure gravity
theory constructed in \cite{Bergshoeff:2010mf}. Thus we shall just
present the result here, but specializing to our specific choice of
parameters.  For those interested in a detailed analysis, we refer
to a recent paper \cite{Bergshoeff:2010iy}. In the case of
$\gamma\equiv \sigma - 1/(2\nu^2)\ne 0$, the transverse traceless
massive graviton modes satisfy
\begin{equation}\
D(\eta_{\pm})h_{\mu\nu}=0\,,\qquad
D(\eta)_{\mu}^{~\nu}=\delta_{\mu}^{~\nu}+
\eta\varepsilon_{\mu}^{~\alpha\nu}\bar{\nabla}_{\alpha}\,,
\label{dheom}
\end{equation}
with
\begin{equation}
\eta_\pm = \gamma^{-1} \Big(\fft{1}{2\mu} \pm \sqrt{\fft{1}{4\mu^2}
- \fft{\gamma}{\nu^2}}\Big)\,.\label{etapm}
\end{equation}
Note that in the limit of $\mu^2\rightarrow \infty$, corresponding
to turning off the Lorentz-Chern-Simons term, we need require that
$\gamma/\nu^2<0$.  In the limit of $\nu^2\rightarrow \infty$, we
have
\begin{equation}
\eta_\pm = \fft{1 + \mu/|\mu|}{2\mu\sigma} = 0\,,\qquad {\rm
or}\qquad \fft{1}{\sigma\mu}\,.\label{etapmspec}
\end{equation}
As we shall see later, the mode with $\eta=0$ becomes infinitely
massive and decouples from the spectrum. The unitarity of the dual
CFT requires that
\begin{equation}
|\eta_{\pm}|\leq1\,.\label{cftunitarity}
\end{equation}
The theory becomes critical when we have $|\eta_+|=1$ or
$|\eta_-|=1$. This can be achieved by
\begin{equation}
\sigma = -\fft{1}{2\nu^2} + \fft{1}{\mu}\qquad {\rm or}\qquad \sigma
= -\fft{1}{2\nu^2} - \fft{1}{\mu}\,.
\end{equation}
The central charges for the right-handed and left-handed Virasoro
algebra of the boundary CFT can be obtained
\cite{Henningson:1998gx}-\cite{Tachikawa:2006sz}, given by
\begin{equation}
C_{L, R}=\frac{3}{2G_3}(\sigma+\frac{1}{2\nu^2} \mp
\frac{1}{\mu})\,.\label{clr}
\end{equation}
Thus we see that the either $C_L$ or $C_R$ vanishes at the critical
points.  To be specific, we summarize the four critical points as
follows:
\begin{eqnarray}
\hbox{case 1}:&&\sigma=\fft{1}{\mu} - \fft{1}{2\nu^2}\,,\qquad
\mu>0\,,\qquad \nu^2\ge 2\mu\,,\cr 
&&\eta_+=1\,,\qquad 0\le\eta_-=\fft{\mu}{\nu^2-\mu}\le 1\,, \qquad
C_L=0\,,\qquad C_R=\fft{2}{\mu}\,;\cr 
\hbox{case 2}:&&\sigma=\fft{1}{\mu} - \fft{1}{2\nu^2}\,,\qquad
\mu>0\,,\qquad \nu^2<0\,, \cr 
&&\eta_+=1\,,\qquad -1<\eta_-=\fft{\mu}{\nu^2-\mu}\le0\,, \qquad
C_L=0\,,\qquad C_R=\fft{2}{\mu}\,;\cr 
\hbox{case 3}:&&\sigma=-\fft{1}{\mu} - \fft{1}{2\nu^2}\,,\qquad
\mu<0\,,\qquad \nu^2<0\,, \cr 
&&0\le\eta_+=\fft{\mu}{\nu^2+\mu}<1\,,\qquad \eta_-=-1\,,\qquad
C_L=-\fft{2}{\mu} \,,\qquad C_R=0\,,\cr 
\hbox{case 4}:&&\sigma=-\fft{1}{\mu} - \fft{1}{2\nu^2}\,,\qquad
\mu<0\,,\qquad \nu^2\ge -2\mu\,, \label{4cases}\\ 
&&-1\le \eta_+=\fft{\mu}{\nu^2+\mu}\le0\,,\qquad \eta_-=-1\,,\qquad
C_L=-\fft{2}{\mu} \,,\qquad C_R=0\,.\nonumber
\end{eqnarray}
In all the above four critical points, the Breitlohner-Freedman
condition (\ref{bfcon}) for the scalar modes and the CFT unitarity
conditions (\ref{cftunitarity}) are satisfied.  Furthermore, the
$\sigma$ in all these cases are positive definite implying that the
matter scalar $\phi$ is ghost free.

It is clear that at the critical points, one of the massive graviton
modes, corresponding to either $\eta_+=1$ or $\eta_-=-1$ and with
vanishing associated central charge, becomes pure gauge and
decouples from the bulk physics.  This feature is the same as chiral
gravity \cite{lisost}.  However, there is an important difference
that a non-trivial massive graviton mode still survives in our
theory at the critical points. It is thus necessary to verify that
this mode has positive energy. To compute the energy of the pure
graviton mode, we set $\phi= s=0$, and also $h=0$. We obtain the
quadratic action for the transverse traceless graviton. After
integrating by parts, we have
\begin{eqnarray}
I^{(2)}&=&\frac{1}{2\kappa^2}\int
d^3x\sqrt{-\bar{g}}\{\frac{1}{2\nu^2}\bar{\nabla}^2
h^{\mu\nu}\bar{\nabla}^2 h_{\mu\nu}-\ft12
(\sigma+\frac{9m^2}{2\nu^2})
\bar{\nabla}^{\lambda}h^{\mu\nu}\bar{\nabla}_{ \lambda}h_{\mu\nu}\cr
&& +(\sigma+\frac{5m^2}{2\nu^2})m^2h^{\mu\nu}h_{\mu\nu}
-\frac{1}{\mu}\varepsilon_{\mu}^{~\alpha\beta}(
\ft12\bar{\nabla}_{\alpha}h^{\mu\nu}\bar{\nabla}^2
h_{\beta\nu}+m^2\bar{\nabla}_{\alpha}h^{\mu\nu} h_{\beta\nu})\}
\end{eqnarray}
Note that we restore the parameter $m$ to keep track the dimensions
of various terms. In the global coordinates, the metric of AdS$_3$
vacuum takes the form
\begin{equation}
ds^2=l^2(-\cosh^2\rho d\tau^2+\sinh^2\rho d\phi^2+d\rho^2)
\end{equation}
where the radius is related to $m$ by
\begin{equation}
    l^2=m^{-2}.
\end{equation}
It is clear that in this coordinate system, the above action has
$\tau$ translational invariance, the corresponding Noether charge is
given by
\begin{equation}
H=\int d^2x(\pi^{(1)\mu\nu}\dot{h}_{\mu\nu}+
\pi^{(2)\mu\nu}\bar{\nabla}_0\dot{h}_{\mu\nu}-L^{(2)})\,,
\end{equation}
where
\begin{eqnarray}
\pi^{(1)\mu\nu} &=& \frac{\sqrt{-\bar{g}}}{
2\kappa^2}[\bar{\nabla}^0(-(\sigma+\frac{9m^2}{2\nu^2})h^{\mu\nu}+
 \frac{1}{2\mu}\varepsilon^{\beta\alpha(\mu}
 \bar{\nabla}_{\alpha}h^{\nu)}_{~\beta}-\nu^{-2}
 \bar{\nabla}^2h^{\mu\nu})\cr
&&-\frac{1}{\mu}\varepsilon^{0\beta(\mu}(\ft12
\bar{\nabla}^2h^{\nu)}_{~\beta}+m^2h^{\nu)}_{\beta})]\,,\cr
\pi^{(2)\mu\nu} &=&\frac{\sqrt{-\bar{g}}}{2\kappa^2}
 g^{00}[\nu^{-2}\bar{\nabla}^2
 h^{\mu\nu}-\frac{1}{2\mu}\varepsilon^{\beta\alpha(\mu}
 \bar{\nabla}_{\alpha}h^{\nu)}_{~\beta}]\,.
\end{eqnarray}
Following \cite{lisost}, we identify the conserved charge as the
energy of the graviton. Using equations of motion, we find that the
above quantities for massive gravitons become
\begin{eqnarray}
\pi^{(1)\mu\nu}_{M}&=&-\frac{\sqrt{-\bar{g}}}{2\kappa^2}
[(2m^2\nu^{-2}+\frac{1}{2\mu\eta_{\pm}})\bar{\nabla}^0h^{\mu\nu}_{M}
+\frac{\nu^2}{2\mu}(\sigma+\frac{m^2}{2\nu^2}-
\frac{1}{\mu\eta_{\pm}})\varepsilon_{\beta}^{~0\mu}h^{\beta\nu}_{M}]
\,,\cr
\pi^{(2)\mu\nu}_{M}&=&-\frac{\sqrt{-\bar{g}}}{2\kappa^2}g^{00}
(\sigma+\frac{5m^2}{2\nu^2}-\frac{1}{2\mu\eta_{\pm}})h^{\mu\nu}_{M}\,.
\end{eqnarray}
Specializing to the linearized graviton, upon utilizing their
equations of motion, we have the energies
\begin{eqnarray}
H_{M} &=& (\sigma+\frac{m^2}{2\nu^2}-\frac{1}{\mu\eta_{\pm}})\int
d^2x\frac{\sqrt{-\bar{g}}}{2\kappa^2}(\bar{\nabla}^0h^{\mu\nu}_{M}
\dot{h}_{M\mu\nu}-\frac{\nu^2}{2\mu}\varepsilon_{\beta}^{~0
\mu}h_{M}^{\beta\nu}\dot{h}_{M\mu\nu})\,.\label{energyres}
\end{eqnarray}
In the limit $\mu\rightarrow \infty$, the energy formula reduces to
the one obtained in \cite{Liu:2009bk} for new massive gravity. The
limit $\nu\rightarrow \infty$ is more subtle.  As we see from
(\ref{etapm}) and (\ref{etapmspec}), one of the $\eta$'s vanishes in
this limit, the corresponding energy (\ref{energyres}) is infinite
and hence its mode decouples from the spectrum.  For
$\eta=1/(\sigma\mu)$, we need expand $\eta$ to the next order in
$1/\nu^2$, and the resulting energy is then precisely the one given
in \cite{lisost}.

The solutions for (\ref{dheom}) were obtained in \cite{lisost}. The
relevant one corresponding to the primary state is given by the real
or imaginary part of $\psi_{\mu\nu}$, where
\begin{equation}
\psi_{\mu\nu}=\fft{e^{-{\rm i} h u - {\rm i} \bar h v}\sinh^2
\rho}{(\cosh\rho)^{h+\bar h}} \left(
\begin{matrix} 1 & \ft12 (h-\bar h) & \fft{2{\rm i}}{\sinh 2\rho}\cr
\ft12(h-\bar h) & 1 & \fft{{\rm i} (h-\bar h)}{\sinh 2\rho} \cr 
\fft{2{\rm i}}{\sinh 2\rho} &  \fft{{\rm i} (h-\bar h)}{\sinh 2\rho}
&-\fft{4}{\sinh^2 2\rho} \end{matrix}\right)\,.
\end{equation}
Here $u=\tau + \phi$ and $v=\tau-\phi$ are the light-cone
coordinates. The solution is parameterized by the weights $(h,\bar
h)$ of the left and right Virasoro algebra of the dual conformal
field theory. The weights are given by \cite{lisost}
\begin{equation}
h=\fft32 \pm \fft{1}{2\eta}\,,\qquad \bar h = -\fft12 \pm
\fft{1}{2\eta}\,.
\end{equation}
For $\eta\in(0,1]$, the plus sign is chosen; for $\eta\in [-1,0)$,
the minus sign is chosen. Note that we have set again $m=1$. At the
four critical points discussed earlier, one of the two massive
graviton becomes pure gauge, and its energy defined by
(\ref{energyres}) indeed vanishes identically.  The energy for the
remaining massive graviton, cataloged as in (\ref{4cases}), is given
by
\begin{eqnarray}
\hbox{Case 1}:&& H_M(\eta_-)\propto \fft{2\nu^2(\nu^2-2\mu)^2(\mu+
\nu^2)}{\mu^3(2\nu^2 -\mu)(\nu^2-\mu)}\,,\cr 
\hbox{Case 2}:&& H_M(\eta_-)\propto
\fft{2(\nu^2-2\mu)^2(\nu^4-\nu^2\mu + 2\mu^2)}{\mu^3(3\mu -
2\nu^2)(\mu-\nu^2)}\,,\cr 
\hbox{Case 3}:&& H_M(\eta_+)\propto
\fft{2(\nu^2-\mu)(\nu^2+2\mu)^3}{\mu^3(\nu^2+\mu)(3\mu+2\nu^2)}\,,\cr
\hbox{Case 4}:&& H_M(\eta_+)\propto
\fft{2\nu^2(\nu^2+2\mu)(\nu^4+\nu^2\mu +
2\mu^2)}{\mu^3(\nu^2+\mu)(\mu+2\nu^2)}\,.
\end{eqnarray}
Examining the range of the parameters $(\mu,\nu^2)$ listed in
(\ref{4cases}), we conclude that the energy for both case 1 and case
2 are non-negative.  For the energy of case 3 to be non-negative, a
condition $\nu^2\ge \mu$ must be further imposed.  The energy of
case 4 is always negative.

Since the analysis of the tensorial modes is the same as that for
pure massive supergravities \cite{Bergshoeff:2010mf,lps}, our results also apply to the pure gravity theory (\ref{newlag1}), for which the trace mode $h$ decouples. Although critical points were obtained previously by examining the linearized equations of motion for pure massive supergravity \cite{Bergshoeff:2010mf}, our results provide further details of the on-shell energy for the surviving massive graviton.

     Thus we have shown that there exist critical points of the
parameters such that one massive graviton becomes pure gauge whilst
the other has positive energy and hence the theory is ghost free.
This strongly suggests that these critical theories, with or without the matter scalar multiplet, may be well-defined at the full quantum level.

\section{Positivity of the BTZ black hole mass}

In the previous section, we demonstrate that the hybrid theory with
five parameters obtained in section 2 has critical points for which
one of the two massive graviton modes becomes trivial and the other
has positive energy.  In this section, we investigate the energy of
the BTZ black hole that is asymptotic to the supersymmetric AdS$_3$
vacuum. The procedure of calculating the mass and angular momentum
of such a black hole in a theory with higher-derivative curvature
terms was spelled out in detail in \cite{deserenergy}. The
modifications to the energy and angular momentum of the BTZ black
hole due to the Lorentz Chern-Simons term were obtained in
\cite{ost}. Here we shall present the formalism of the modifications
due to the $\nu$ term.  For our purpose, we set $\phi=0$ and $S=-m$.
We expand the BTZ black hole around the AdS$_3$ background as
$g_{\mu\nu} = \bar g_{\mu\nu} + h_{\mu\nu}$, where
\begin{equation}
\bar{R}_{\mu\nu\alpha\beta}=-\bar{g}_{\mu\alpha}\bar{g}_{\nu\beta}+
\bar{g}_{\mu\beta}\bar{g}_{\nu\alpha}\,,\qquad
\bar{R}_{\mu\nu}=-2\bar{g}_{\mu\nu}\,,\qquad \bar{R}=-6\,.
\end{equation}
Note that we have set $m=1$ for convenience. Expressing the Einstein
equations of motion as $E_{\mu\nu}=0$, we can expand the equations
around the background, giving
\begin{equation}
E^{L}_{\mu\nu}\equiv\kappa^2 T_{\mu\nu}\,,
\end{equation}
where $L$ denotes the linear part of the Einstein equations. All the
non-linear quantities are lumped together and expressed as
$T^{\mu\nu}$. Because of the linearized Bianchi indentity
$\bar{\nabla}_{\mu}E^{L\mu\nu}=0$, $T_{\mu\nu}$ is covariantly
conserved with respect to the background metric. Therefore
$j^{\mu}\equiv T^{\mu\nu}\xi_{\nu}$ is a conserved current, where
$\xi^\mu$ is a Killing vector in the background metric. The
corresponding conserved charge is given by
\begin{equation}
    Q(\xi)=\fft{1}{8\pi G_3}\int d\Sigma_{\mu}j^{\mu}\,.
\end{equation}
Here $\Sigma$ is a two-dimensional space-like hypersurface and
$d\Sigma_{\mu}= n_{\mu}\sqrt{\gamma}d^2x$ where $n_{\mu}$ is the
unit normal vector and $\gamma$ is the determinant of the induced
metric on $\Sigma$.

In our case with the $\sigma$ and $\nu$ terms, we have
\begin{eqnarray}
&&\kappa^2 T_{\mu\nu}=
E^{L}_{\mu\nu}=(\sigma+\frac{3}{2\nu^2})\mathcal {G}^{L}_{\mu\nu}
+ \frac{1}{4\nu^2}(\bar{g}_{\mu\nu}\bar {\Box}-\bar{\nabla}_{\mu}
\bar{\nabla}_{\nu}-2\bar{g}_{\mu\nu})R^{L}\\
&&+\frac{1}{\nu^2}(\bar{\Box}\mathcal
{G}^{L}_{\mu\nu}+\bar{g}_{\mu\nu}R^{L})
 +\frac{1}{\mu\sqrt{-\bar{g}}}\epsilon^{\mu\alpha\beta}
\bar{g}_{\beta\sigma} \bar{\nabla}_{\alpha} \left( R^{\sigma\nu}_{L}
- 2 \Lambda h^{\sigma\nu} - \frac{1}{4} \bar{g}^{\sigma\nu}
R_{L} \right)\,,
\end{eqnarray}
where
\begin{eqnarray}
R^L &=& -\bar \Box h + \bar\nabla_\mu\bar\nabla_\nu h^{\mu\nu} + 2
h\,,\qquad h = \bar g^{\mu\nu} h_{\mu\nu}\,,\cr 
R^L_{\mu\nu} &=& \ft12(-\bar\Box h_{\mu\nu} - \bar \nabla_\mu \bar
\nabla_\nu h + \bar \nabla^\rho \bar \nabla_\nu h_{\rho\mu} + \bar
\nabla^\rho \bar \nabla_\mu h_{\rho\nu})\,,\cr 
{\cal G}^L_{\mu\nu} &=& R^L_{\mu\nu} - \ft12 \bar g_{\mu\nu} R^L + 2
h_{\mu\nu}\,.
\end{eqnarray}
We find that  $j^{\mu}$ can be expressed as
$\nabla_{\nu}\mathcal{F}^{\mu\nu}$, where
\begin{eqnarray}
\mathcal {F}^{\mu\nu} &=& (\sigma+\frac{1}{2\nu^2})\{
\xi_{\alpha}\bar{\nabla}^{\mu}h^{\nu\alpha}-
\xi_{\alpha}\bar{\nabla}^{\nu}h^{\mu\alpha}
+\xi^{\mu}\bar{\nabla}^{\nu}h-\xi^{\nu}\bar{\nabla}^{\mu}h
+h^{\mu\alpha}\bar{\nabla}^{\nu}\xi_{\alpha}-h^{\nu\alpha}
\bar{\nabla}^{\mu}\xi_{\alpha}\cr 
&&+\xi^{\nu}\bar{\nabla}_{\alpha}h^{\mu\alpha}-
\xi^{\mu}\bar{\nabla}_{\alpha}h^{\nu\alpha}
+h\bar{\nabla}^{\mu}\xi^{\nu}\}+\frac{1}{4\nu^2}
\{\xi^{\mu}\bar{\nabla}^{\nu}R^{L}-\xi^{\nu}
\bar{\nabla}^{\mu}R^{L}+R_{L}\bar{\nabla}^{\mu}\xi^{\nu}\}\cr
&&+\frac{1}{\nu^2}\{\xi_{\alpha}\bar{\nabla}^{\nu}\mathcal
{G}^{L\mu\alpha}-\xi_{\alpha}\bar{\nabla}^{\mu}\mathcal
{G}^{L\nu\alpha}-\mathcal
{G}^{L\mu\alpha}\bar{\nabla}^{\nu}\xi_{\alpha}
   + \mathcal {G}^{L\nu\alpha}\bar{\nabla}^{\mu}\xi_{\alpha}
   \}+\mathcal{F}^{\mu\nu}(\mu)\,,\label{fmunu}
\end{eqnarray}
where the $\mu$-term standing for contribution from Chern-Simons term can be
read from the formulae presented in \cite{ost}.
Using Stokes theorem the conserved charge can now be expressed as
\begin{equation}
    Q(\xi)=\fft{1}{8\pi G_3}\int\mathcal {F}^{\mu\nu}dS_{\mu\nu}\,,
\end{equation}
where $S$ is the boundary of $\Sigma$.

We now examine a specific example, namely the BTZ black hole that is
asymptotic to the supersymmetric AdS$_3$ vacuum.  The solution is
given by \cite{btz}
\begin{eqnarray}
ds^2=-U dt^2 + \fft{dr^2}{U} + r^2 (d\phi -
\fft{4J}{r^2}dt)^2\,,\qquad U=r^2 - 8M + \fft{16J^2}{r^2}\,.
\end{eqnarray}
From now on we set $G_3$ to be 1. One may define the mass $E$ and angular momentum $L$ as the
conserved charges associated with the Killing vectors
$K_t=\partial/\partial_t$ and $K_\phi=\partial/\partial_\phi$
respectively. It is straightforward to work out these quantities,
given by
\begin{equation}
E=\Big(\sigma + \fft{1}{2\nu^2}\Big) M - \fft{J}{\mu}\,,\qquad
L=\Big(\sigma + \fft{1}{2\nu^2}\Big) J -
\fft{M}{\mu}\,.\label{btzel}
\end{equation}
Note that the second and third brackets in (\ref{fmunu}) converges
too fast to give any contributions.

     To validate the above mass and angular momentum, we examine
the first law of thermodynamics.  The temperature and the angular
velocity can be obtained directly from the metric, given by
\begin{equation}
T=\fft{r_+^4-16J^2}{2\pi r_+^3}\,,\qquad \Omega=\fft{4J}{r_+^2}\,,
\end{equation}
where
\begin{equation}
r_\pm = \sqrt{2(M+J)} \pm \sqrt{2(M-J)}\,.
\end{equation}
The entropy can be obtained using the Cardy formula {\it via} the
AdS/CFT correspondence, given by \cite{lisost}
\begin{equation}
S=\ft13\pi^2 C_L T_L + \ft13 \pi^2 C_R T_R\,,
\end{equation}
where the central charges $C_{L,R}$ are given by (\ref{clr}) and
$T_{L,R}$ are give by
\begin{equation}
T_L=\fft{r_+-r_-}{2\pi}\,,\qquad T_R=\fft{r_+ + r_-}{2\pi}\,.
\end{equation}
It is now straightforward to verify the first law of thermodynamics,
namely
\begin{equation}
dE = TdS + \Omega dL\,.
\end{equation}
The result should not be surprising since the effect of $\nu$ in
thermodynamical quantities is to shift the parameter $\sigma$
uniformly.

  For the BTZ black hole, we have $M\ge |J|$; nevertheless, the mass
can be negative for generic parameters $(\sigma, \nu^2, \mu)$.
However, recall that the parameter conditions for the critical
points (\ref{4cases}), we have
\begin{equation}
E_{\rm crit} = \fft{M}{|\mu|} - \fft{J}{\mu}\,,\qquad L_{\rm crit} =
\fft{J}{|\mu|} - \fft{M}{\mu}\,.
\end{equation}
Thus we see that at the critical points, the mass is non-negative.
Furthermore the quantity $(E_{\rm crit}-L_{\rm crit})$ is either
positive for $\mu>0$ or 0 for $\mu<0$, but never negative.

It is worth pointing out that since the BTZ black hole does not involve the scalar multiplet, its mass calculation is the same for that in pure massive supergravity.  Our results thus demonstrate the negativeness of the mass of the BTZ black for pure massive supergravity, which was not studied previously.  The effect of adding scalar multiplet provides an additional constraint to the parameters.  This can be seen more clearly by considering the case where the Lorentz-Chern-Simons term is turned off, by setting $\mu\rightarrow \infty$. In this case at the critical point, there is no massive graviton and the BTZ black hole has zero energy and angular momentum.  Thus for pure gravity, the sign choice of $\sigma$ can be either positive of negative.  The inclusion of the scalar multiplet will force that $\sigma$ to be positive.

\section{BPS string solution}

\subsection{The solution}

In this section, we construct BPS string solutions with $R^{1,1}$
isometry. The ansatz is given by
\begin{equation}\label{stringans}
    ds^2=dz^2+e^{2A(z)}(-dt^2+dx^2)\,,\qquad \phi=\phi(z)\,,\qquad
    S=S(z)\,.
\end{equation}
(Such a metric ansatz is also called domain wall.) It is worth
pointing out immediately that for this metric ansatz, the ${\cal
L}_{\rm LCS}$ term gives no contribution to the equations of motion.
A natural choice for the vielbein is given by
\begin{equation}
    e^0=e^{A}dt\,,\qquad e^1=e^Adx\,,\qquad e^2=dz\,.
\end{equation}
The non-vanishing components of the spin connection and curvature
are
\begin{eqnarray}
&& \omega_{02}=-A^{'}e^0,,\qquad \omega_{12}=A^{'}e^1\,,\cr
&& R_{0101}=A^{'2}\,,\qquad R_{0202}=A^{''}+A^{'2}\,,\qquad
R_{1212}=-R_{0202},.\label{concurv}
\end{eqnarray}
Here a prime denotes a derivative with respect to $z$.  The
three-dimensional gamma matrices can be chosen to be
\begin{equation}
    \gamma_0=\left(
               \begin{array}{cc}
                 0 & 1\\
                 -1 & 0 \\
               \end{array}
             \right),\qquad
    \gamma_1=\left(
               \begin{array}{cc}
                 0 & 1 \\
                 1 & 0 \\
               \end{array}
             \right),\qquad
    \gamma_2=\left(
               \begin{array}{cc}
                 1 & 0 \\
                 0 & -1 \\
               \end{array}
             \right).
\end{equation}
It follows from (\ref{susytrans}) that a Killing spinor $\epsilon$
satisfies
\begin{eqnarray}
D_{\mu}\epsilon+\frac{1}{2}\gamma_{\mu}S\epsilon &=&
0\,,\label{ks1}\\
(\gamma^{\mu}\partial_{\mu}\phi+S+m)\epsilon &=& 0\,.\label{ks2}
\end{eqnarray}
From (\ref{ks2}), we deduce that
\begin{equation}
    (\partial\phi)^2=(S+m)^2,
\end{equation}
Thus for our string ansatz, we have, without loss of generality,
\begin{equation}
    \phi^{'}=-(S+m),
\end{equation}
This corresponds to the following projection
\begin{equation}\label{ksproj}
    (\gamma_2-1)\epsilon=0\qquad\Rightarrow
\qquad \epsilon=\left(
\begin{array}{c}\chi \\0 \\\end{array}\right).
\end{equation}
The integrability condition for (\ref{ks1}) is
\begin{equation}
\ft14 R_{\alpha\beta\nu\mu}\gamma^{\alpha\beta}\epsilon+ \ft12
(\gamma_{\mu}\partial_{\nu}S-\gamma_{\nu}
\partial_{\mu}S)\epsilon+\ft12\gamma_{\nu\mu}S^2\epsilon=0.
\end{equation}
Substituting the Riemann curvature (\ref{concurv}) and the killing
spinor (\ref{ksproj}), we find that the only solution is
\begin{equation}\label{sarelation}
    S=-A^{'}\,.
\end{equation}
The Killing spinor can then be solved explicitly from (\ref{ks1}),
given by
\begin{equation}\label{ksres1}
\epsilon=\left(\begin{array}{c} e^{\frac{1}{2}A}\\
0\\\end{array} \right)
\end{equation}
Under the string ansatz, the $\phi$ equation (\ref{phieom}) becomes
\begin{equation}
(8m+2S+6A^{'})(A^{'}+S)+4(S+A^{'})^{'}=0\,.
\end{equation}
It is automatically satisfied by (\ref{sarelation}). The $S$
equation (\ref{seom}) can be simplified and becomes
\begin{equation}
(\alpha + 8 b) S'' +2(4b+3c) SS' - 2a S' + c S^3 + \sigma e^{-2\phi}
S + \tilde m=0\,.\label{stringseom}
\end{equation}
The Einstein equations of motion (\ref{einstein}) are then all
automatically satisfied.

It is now straightforward to obtain the supersymmetric AdS vacuum
solution, given by
\begin{equation}
\sigma=\fft{\tilde m - c m^3}{m}\,,\qquad S=-m\,,\qquad
\phi=0\,,\qquad R_{ij}=-2m^2 g_{ij}\,.
\end{equation}
Here we let $\sigma$ be continuous so that $\phi$ is zero instead of
a non-vanishing constant. The general solution for
(\ref{stringseom}) is not expected to be solved explicitly.  For
certain choices of the parameters, explicit solutions can be
obtained.  Let us first consider the case with
\begin{equation}
(\alpha,a,b,c)=0\,.
\end{equation}
The solution is given by
\begin{eqnarray}
ds^2 &=& \fft{e^{2m z}}{1 + q e^{2m z}} (-dt^2 + dx^2) +
dz^2\,,\cr
e^{-2\phi} &=& \fft{\tilde m}{m} (1 + q e^{2m z})\,,\qquad
S=-\fft{m}{1 + q e^{2m r}}\,.
\end{eqnarray}
The coordinate $z$ runs from $-\infty$ to $+\infty$, and the metric
interpolates between the AdS$_3$ horizon and asymptotic flat
Minkowski space-time.  The string coupling constant $g=e^{\phi}$
runs from the constant $\sqrt{m/\tilde m}$ to $0$ at the asymptotic
flat region. The solution can be lifted to $D=6$ and it becomes a
dyonic string with the ``1'' in the harmonic function associated
with the magnetic component dropped.  To be specific, we have
\begin{eqnarray}
ds^2_6 &=& H_e^{-1} (-dt^2 + dx^2) + H_m (dr^2 + r^2
d\Omega_3^2)\,,\cr 
e^{-2\phi} &=& \fft{\tilde m}{m} \fft{H_m}{H_e}\,,\qquad H_\3 =
dx\wedge dt \wedge dH_e^{-1} + m^2 \Omega_\3\,,\cr 
H_e &=& 1 + \fft{q}{r^2}\,,\qquad H_m =\fft{m^2}{r^2}\,.
\end{eqnarray}

   The second case we would like to consider is the following
\begin{equation}
a=0\,,\qquad b=-\ft34 c\,,\qquad \alpha = 6 c\,,\qquad \tilde m=0\,.
\end{equation}
We find the solution is given by
\begin{eqnarray}
ds^2 &=& (e^{m z}-q)^2 (-dt^2 + dx^2) + dz^2\,,\cr 
S&=& - \fft{m}{1 - q e^{- m z}}\,,\qquad e^{-2\phi} = -\fft{c m}{(1
- q e^{-m z})^2}\,.\label{string1}
\end{eqnarray}
In this case the metric approaches AdS$_3$ at the asymptotic
$z\rightarrow \infty$ and has a naked singularity in the middle.

 The above two solutions can be grouped together with the parameter
choice (\ref{auxcon}). In this case, it is advantageous to make a
coordinate transformation and treat $S$ as the radial coordinate.
The solution is then given by
\begin{eqnarray}
ds^2&=& e^{2A} (-dt^2 + dx^2) + \fft{dS^2}{S^2 f^2} \,,\cr 
e^{2\phi} &=& \fft{S}{S^3-S_2^3}\,,\qquad A=\int \fft{1}{f} dS
\qquad f=-\fft{2(S-S_1)(S^3-S_2^3)}{S_2^3+ 2S^3}\,,\label{solution2}
\end{eqnarray}
where
\begin{equation}
S_1=-m <0\,,\qquad S_2=-(6\nu^2\tilde m)^{1/3}<0\,.
\end{equation}
The explicit expression for $A$ is somewhat complicated, given by
\begin{eqnarray}
A&=&A_0 + \fft{1}{4(S_1^3-S_2^3)}\Big(-2\sqrt3
S_1S_2(S_1-S_2)\arctan(\fft{2S+S_2}{\sqrt3 S_2})\cr
&&\qquad -2(2S_1^3+S_2^3) {\rm log}(S-S_1) +2 S_1S_2(S_1 + S_2){\rm
log}(S-S_2)\cr 
&&\qquad -S_1S_2(S_1 + S_2){\rm log}(S^2+S S_2 + S_2^2) + 2S_2^3
{\rm log} (S^3-S_2^3)\Big)\,,
\end{eqnarray}
where $A_0$ is an integration constant and it should be chosen
appropriately such that the expression for $A$ is real.

In the vicinity of $S=0$, the metric describes Minkowski space-time.
In the vicinity of $S=S_1\equiv -m<0$, the solution approaches the
vacuum AdS$_3$.  In the vicinity of $S=S_2\equiv -(\tilde
m/c)^{1/3}<0$, the solution becomes
\begin{equation}
ds^2 \sim e^{-2S_2 z} (-dt^2 + dx^2) + dz^2\,,\qquad \phi\sim
(S_1-S_2) z\,.
\end{equation}
This linear-dilaton solution is approximate, valid for $z\rightarrow
\infty$ when $S_2<S_1$, and for $z\rightarrow -\infty$ when
$S_2>S_1$. For both cases, the ``string'' coupling $g=e^{\phi}$ goes
to infinity.  It is clear that in the region $S\in({\rm
max}(S_1,S_2),0)$, the metric interpolates between the flat
Minkowski space-time and the horizon of an AdS$_3$.  For $S_1 <
S_2<0$, the region of $S\in (S_1,S_2)$ describes an interpolation
between the boundary of the vacuum AdS$_3$ and the AdS$_3$ with the
linear dilaton. To see this, we note that the function $f$ is
negative in this region and that
\begin{equation}
A=\int^S_{S_0} \fft{1}{f} dS\,,
\end{equation}
where $S_1<S_0<S_2$.  Thus $A(S_1)\rightarrow +\infty$ and
$A(S_2)\rightarrow -\infty$, indicating a boundary and a horizon
structure respectively.   For $S_2<S_1<0$, the role the two AdS$_3$
reverses, and the horizon lies in the vacuum AdS$_3$ whilst the
boundary lies in the AdS$_3$ with the linear dilaton.

\subsection{Spectrum analysis}

In the previous subsection, we discuss the general BPS string
solutions.  For certain parameter choice, we obtain explicit
solutions.  Some of these metrics are asymptotic to AdS$_3$ and
hence are dual to certain two-dimensional field theory that has an
ultra-violet conformal fix points.  In the general discussion of the
AdS/CFT correspondence, the correlation functions and the spectrum
of the corresponding strongly coupled two-dimensional theory can be
analyzed by studying the wave equations in these gravitational
backgrounds in the Einstein frame. Making appropriate coordinate
transformations, the metric can be cast into the following form.
\begin{equation}
ds^2=e^{2\tilde A} (-dt^2 + dx^2 + dr^2)\,,\label{conformalform}
\end{equation}
The simplest two-point function is that of the operator ${\cal
O}\sim \tr F^2$, which is expected to be coupled to a massless
$s$-wave free scalar $\phi$, satisfying
\begin{equation}
\partial_{\mu} (\sqrt{-g} g^{\mu\nu} \partial_\nu \phi)=0\,.
\end{equation}
For the $s$-wave, we have $\phi=e^{i\omega t} e^{-\fft12\tilde A}
\chi(r)$, where $\omega$ measures the energy level of the solution.
It is easy to show that $\chi$ satisfies Schr\"odinger equation
\begin{equation}
\Big[-\partial_r^2 + V(r)\Big]\chi = \omega^2 \chi\,,
\end{equation}
where
\begin{equation}
V=\ft14 (2\tilde A'' + \tilde A'^2)\,.
\end{equation}
Note that the potential can be written as $V=U^2+U'$ where the
superpotential is $U=\frac{1}{2}A'$, thus this is a supersymmetric
quantum mechanics system \cite{Deger:2002hv}. We now consider the
solution (\ref{string1}). First we convert the metric to the
Einstein frame by a conformal rescaling of the metric
$ds^2\rightarrow e^{4\phi} ds^2$.  We can then cast the resulting
metric into the conformal form (\ref{conformalform}), with
\begin{equation}
e^{2\tilde A} = \fft{e^{6mq r}}{(1-e^{mqr})^2}\,.
\end{equation}
Thus the potential $V$ is
\begin{equation}
V=\fft{m^2q^2 (4e^{2mqr} -10 e^{mqr} + 9)}{4(1-e^{mqr})^2} \,.
\end{equation}
The variable $r$ runs from $-\infty$ to $0$.  In this region, we
have $\ft49 m^2 q^2 \le V <\infty$.  Thus, it is clear that the
spectrum is continuous with a mass gap $V_{\rm min}=\ft94 m^2 q^2$.

     The structure for the solution (\ref{solution2}) is more
complicated, and the explicit form of the potential $V$ cannot be
obtained.  The characteristics of the potential can nevertheless be
analyzed.  Although in the string frame, the metric runs from an
AdS$_3$ horizon to an AdS$_3$ boundary, in the Einstein frame, one
of the AdS$_3$ turns into a singular metric owing to the linear
dilaton.  For the theory to dual to a two-dimensional quantum field
theory, it is easier to consider the case with $S_1<S_2$ so that its
asymptotic behavior is AdS$_3$ with a naked singularity in the bulk.
Following the same procedure outlined for the previous simpler
example, we obtain that now $r$ lies in the region $(r_1,r_2)$ with
the potential $V$ behaves as follows
\begin{eqnarray}
r\rightarrow r_2&& V \sim \fft{3}{4(r-r_2)^2}\,, \cr 
r\rightarrow r_1 && V \sim
\fft{(3S_2-2S_1)(5S_2-2S_1)}{4S_2^2(r-r_1)^2}\,.
\end{eqnarray}
Thus in general the system has a discrete spectrum.

\section{General supersymmetric solutions}

By definition, in any supersymmetric background, there exists a
solution of the Killing spinor equations (\ref{ks1}, \ref{ks2}).
From the Killing spinor $\epsilon$, we can construct a Killing
vector
\begin{equation}
    K^{\mu}=\bar{\epsilon}\gamma^{\mu}\epsilon\,,
\end{equation}
which is null.  It follows from (\ref{ks1}) that we have
\begin{equation}\label{gensusyint}
S\,K^{\lambda}=\frac{\epsilon^{\lambda\alpha\beta}}{2\sqrt{-g}}
\partial_\alpha K_\beta,\qquad \epsilon^{012}=1\,.
\end{equation}
As was demonstrated in \cite{gps}, the general metric ansatz with a
null Killing vector $K=\partial/\partial_v$ can be casted into the
following form
\begin{equation}
   ds^2=e^{2A}dz^2+e^{2B}du^2+2e^{2mz}dudv,
\end{equation}
where the functions $A$ and $B$ depend on the coordinates $(u,z)$.
The non-vanishing components of the Ricci tensor are given by
\begin{eqnarray}
R_{uu} &=& -\ft12e^{-2A} [h''-(2m+A')h'+4m^2h]
-(\ddot{A}+\dot{A}^2)\,,\cr 
R_{uv} &=& m(A'-2m)e^{2(mz-A)}\,,\qquad  R_{uz}=m\dot{A}\,,\qquad
R_{zz}=2m(A'-m)\,,
\end{eqnarray}
where a prime and a dot denote a derivative with respect to $z$ and
$u$ respectively and $h=e^{2B}$.  The Ricci scalar is given by
\begin{equation}
R=e^{-2A}(4mA'-6m^2)\,.
\end{equation}

It follows from (\ref{gensusyint}) that we have the following
algebraic constraint
\begin{equation}\label{susyintres}
    S=-me^{-A}.
\end{equation}
Substituting this relation into (\ref{ks1}), we can find that the
illing spinor is given by
\begin{equation}
    \epsilon=\left(\begin{array}{c}e^{-\frac{1}{2}B+mz}\\
0\\\end{array} \right)
\end{equation}
Comparing this to the Killing spinor (\ref{ksres1}) in the earlier
section, one can deduce that when $B=mz$, the general solutions
reduce to the previous simpler string solution.

Since $g^{uu}=0$, it follows from (\ref{ks2}) that we have $e^{-A}
\phi' = \pm(S+m)$.  The $\pm$ sign choices are inequivalent.  The
plus-sign choice leads only to $S=-m$ and constant $\phi$, which we
shall discuss later.  For the minus-sign, {\it i.e.}
\begin{equation}
e^{-A} \phi'=-(S + m)\,,\label{phiprimem}
\end{equation}
it is easy to verify that the dilaton equation (\ref{phieom}) is
satisfied by (\ref{phiprimem}), together with (\ref{susyintres}).
The $S$ equation of motion (\ref{seom}) now becomes
\begin{equation}\label{eq2}
(\alpha + 8b) S(S^2)'' + 4 am\, S S' - 4 (4b+3c)m\, S^2 S' + 2m^2 (c
S^3 + e^{-2\phi} S + \tilde m)=0\,.
\end{equation}
The Einstein equations of motion are generally satisfied by
(\ref{susyintres}, \ref{phiprimem}, \ref{eq2}) except for one in the
$(u,u)$ direction.  This equation is effectively a 4'th-order linear
differential equation, but it is rather complicated and we shall not
present it here.

     It is worth pointing out that the three equations
(\ref{susyintres}, \ref{phiprimem}, \ref{eq2}) are self-containd
involving three functions $(A,\phi,S)$ and derivatives on coordinate
$z$.  Thus these can be viewed as ordinary differential equations
with the integration constant depending on the coordinate $u$. The
Einstein equation then determines the metric function $B(u,z)$.

Let us consider a simple case with $(a,b,c,\alpha)=0$.  The
functions $(A,\phi,S)$ can be solved explicitly, given by
\begin{equation}
e^{-2\phi}=-\fft{\tilde m}{S}\,,\qquad e^{A}=-\fft{m}{S}\,,\qquad
S=-(m + q(u) e^{2m z})\,.
\end{equation}
The Einstein equation implies that
\begin{eqnarray}
&&\beta\,h''' + \fft{1}{2S^2}\Big(6\beta m S(2m+S) -\tilde m\Big)
h'' - \fft{m}{S^2}\Big(2\beta m (3S^2+2mS - 2m^2) - \tilde
m\Big)h'\cr && -\fft{8\beta m^3(m + S)^2}{S^2}h +
\fft{m^2}{S^6}\Big(4\beta m^2 S(S\ddot S - \dot S^2) + \tilde m
(2S\ddot S - 3 \dot S^2)\Big)=0\,,
\end{eqnarray}
where $h(u,z)=e^{2B}$.  This can be viewed as an ordinary linear
differential equation for $h$ with a source, but with integration
constants now being arbitrary functions of $u$.

The general explicit solutions for the special case with $q(u)=0$
was obtained in \cite{gps}.  Alternatively if we set $\beta=0$,
corresponding to turning off the Lorentz Chern-Simons term, the
general explicit solution also exists, given by
\begin{equation}
e^{2B}=h=f_1(u) + e^{2mz} f_2(u) - \fft{\dot S^2}{4(m+S)^2 S^2} +
\Big(\fft{m- 4mz S - 4z S^2}{2m^2(m+S) S} + \fft{\log(-S)}{m^3}\Big)
\ddot S\,.
\end{equation}
Here $f_1$ and $f_2$ are two arbitrary functions of $u$. It is clear
that $f_2$ can be absorbed by a gauge transformation $v\rightarrow
v-\frac{1}{2}\int f_2du$.

We now consider the parameter choice (\ref{auxcon}).  In this case,
we have
\begin{equation}
e^{-2\phi}=-\fft{cS^3 + \tilde m}{S}\,,\qquad e^{A}=-\fft{m}{S}\,,
\end{equation}
where $S$ satisfies
\begin{equation}
S'=f\equiv \fft{2m(S+m)(cS^3 +\tilde m)}{\tilde m - 2 c S^3}\,.
\end{equation}
This can be solved as follows
\begin{equation}
z + y(u) = \int \fft{dS}{f}\,,
\end{equation}
where $y$ is an arbitrary function of $u$.  This implies that
\begin{equation}
dz + dy= \fft{dS}{f}\,,\qquad \dot S=f \dot y\,.
\end{equation}
Thus we may chose $(S,y)$ as coordinates to replace the original
$(z,u)$.  The Einstein equation gives rise to a linear differential
equation with only $z$ derivative on $h=e^{2B}$ up to the 4'th
order.  The detail expression is rather complicated and we shall not
present here.  The result is simplified significantly if we further
set $\tilde m=0$.  In this case, $S$ can be solved explicitly, given
by
\begin{equation}
S=-(m + q(u) e^{-m z})\,,
\end{equation}
where $q$ is an arbitrary function of $u$.  The function $h=e^{2B}$
satisfies the following equation
\begin{eqnarray}
&&c h'''' + \fft{m}{3S}\Big(\beta m - 6c (3m + 5S)\Big)h''' \cr 
&&+\fft{m^2}{3S^2}\Big(-3\beta m(m+2S) + c (21m^2 + 117 m S + 109
S^2)\Big)h'' \cr 
&&+\fft{m^3}{3S^3}\Big(\beta m (m^2 + 11 m S + 12 S^2) - c(3m^3 + 81
m^2 S + 246 mS^2 + 170 S^3)\Big)h'\cr 
&&+ \fft{2m^4(m+S)}{3S^3} \Big(-\beta m(m+4S) + 3c (m^2+12 m S + 16
S^2)\Big)h\\ 
&&+\fft{2m^4}{3S^6}\Big(\beta m^2 (4\dot S^2 - S \ddot S) + c(24 m^2
\dot S^2 - 6 m S\dot S^2 - 3m^2 S \ddot S + 6S^2 \dot S^2 - 2 S^3
\ddot S)\Big)=0\,.\nonumber
\end{eqnarray}
Since there is only $z$ derivatives on $h$, this is effectively an
fourth-order ordinary linear differential equation with a source.
The integration constants should be considered as arbitrary
functions of $u$.

     A special class of pp-wave solution corresponds to setting
\begin{equation}
S=-M\,,\qquad \phi=0\,,\qquad A=0\,.
\end{equation}
The $S$ equation (\ref{seom}) then requires that $\tilde m=m +
cm^3$.  The Einstein equations reduce to the following differential
equation
\begin{eqnarray}
&&\alpha h'''' -2(2\alpha + \beta) m h''' + (1 + 2am + (4\alpha + 6
\beta + 3c) m^2) h''\cr 
&&\qquad -2m (1+2am +(2\beta + 3c)m^2)h'=0\,.
\end{eqnarray}
The solution is given by
\begin{equation}
h=f_1(u) + f_2(u) e^{2mz} + f_3(u) e^{c_+ z} + f_4(u) e^{c_- z}\,,
\end{equation}
where $f_i$'s are four arbitrary functions of $u$ and constants are
given by
\begin{equation}
c_\pm=\alpha^{-1} \Big[(\alpha + \beta)m \pm \sqrt{(\alpha^2 +
\beta^2 - 3\alpha c) m^2 -\alpha(1 +2am)}\Big]\,.
\end{equation}
It is clear that when $\alpha=\beta=c=a=0$, the solution reduces to
the one given in \cite{gps} for the original topologically massive
supergravity.  There are two additional classes of special
solutions. The first corresponds to having $c=-(1+2a m + 2\beta
m^2)/(3m^2)$ such that $c_+=0$ or $c_-=0$.  The function $h$ is
given by
\begin{equation}
h=f_1(u) + f_2(u) e^{2mz} + f_3(u) z + f_4(u)
e^{2mz(1+\beta/\alpha)}\,.
\end{equation}
The second class corresponds to have $c=-(1+2am - 2\alpha
m^2)/(3m^2)$ and $\beta=-\alpha$, for which $c_+=0=c_-$.  The
function $h$ is given by
\begin{equation}
h=f_1(u) + f_2(u) e^{2mz} + f_3(u) z + f_4(u) z^2\,.
\end{equation}
In all of the above solutions, the terms of $f_1$ and $f_2$ are pure
gauge \cite{gps}. Note that the coordinate $z$ is a logarithmic
function of the global radial coordinate. Specializing the
parameters to the theory discussed in section 2, we find that the
first case in the above corresponds to the critical conditions. Thus
logarithmic modes can emerge at the critical points and to
understand them {\it via} a logarithmic CFT \cite{Grumiller:2009sn},
one should choose a boundary condition \cite{Liu:2009pha} that is
less restrictive than that advocated in \cite{Brown:1986nw}.

\section{Conclusions}

In this paper we study generalized topologically massive
supergravity that was recently constructed in \cite{lps}.  The
theory is hybrid in a sense that it consists one off-shell
supergravity multiplet and one on-shell matter scalar multiplet.  An
important feature of three-dimensional massive supergravities is
that the supersymmetry can be realised off shell. As a consequence, such a theory can be complete in terms of supersymmetry by augmenting with
only a finite number of higher-derivative terms. The hybrid theory
studied in this paper implies that the matter sector can still be
minimally dynamical with at most two derivatives even though the
higher derivative terms in gravity sector is inevitable in a quantum
theory. For generic parameters, the auxiliary scalar field in the
off-shell supergravity multiplet is dynamical. We focus on a special
class of parameter choices for which the auxiliary field is
non-dynamical.  The resulting theory has five parameters and it can
be viewed as generalized super NMG theory with a matter scalar
multiplet.

Since the super NMG theory with negative Einstein-Hilbert action can
be ghost free, we analyze the linear perturbation of our generalized
super NMG theory in Minkowski space-time.  We find that the spectrum
contains one massless scalar mode and two massive graviton modes.
For general parameters, when the Einstein-Hilbert action is
negative, the two graviton modes are ghost free, but the scalar mode
is ghost like. When the action is positive, the scalar mode is
non-ghost, but one of the two massive graviton is ghost like.
However, we find a special choice of parameters such that the
on-shell Hamiltonian for both massive graviton vanish. This signals
that the linearized analysis breaks down and it suggests a
possibility that the theory may become ghost free. A proper
higher-order analysis is relegated for future research.

    Next, we perform linearized analysis around the
supersymmetric AdS$_3$ vacuum.  For generic parameters, the theory
contains one scalar mode and two non-trivial massive gravitons.  For
the scalar mode to be ghost free, it is necessary that the
coefficient $\sigma$ of the super invariant involving the
Einstein-Hilbert term is positive.  Then one of the two massive
gravitons becomes inevitably ghost like.  We obtain the critical
conditions for which the ghost graviton becomes pure gauge and
decouples from the bulk physics. We then show that the remaining
massive graviton can indeed have positive energy. (This conclusion
also applies to the corresponding pure massive supergravity
(\ref{newlag1}).) Furthermore, we also demonstrate explicitly that
the scalar mode is stable satisfying the Breitlohner-Freedman bound.
These properties suggest that the theory may be well-defined at
these critical points.  In order to establish this point further, we
obtain the mass and angular momentum for the BTZ black hole that is
asymptotic to the AdS$_3$.  We find that indeed at the critical
points, the mass is non-negative and furthermore it is always
greater or equal to the angular momentum.  We also verify explicitly
that the first law of thermodynamics holds.

   It should be emphasized that although our analysis focused
on the hybrid theory, the linearized analysis of the traceless graviton mode and the BTZ energy calculation apply equally well to the four-parameter pure massive gravity.  The critical points for the general massive pure supergravity were obtained in \cite{Bergshoeff:2010mf}. Our futher examination of the energy of the remaining non-trivial massive graviton and the mass of the BTZ black hole reveals additional properties of the theory at these critical points.  The inclusion of the on-shell matter multiplet does not alter these results, but instead provides further constraints on the parameter space.

  We also construct BPS solutions of the general theory.  We find that
the equations are reduced to effectively two linear differential
equations of two functions.  For some specific parameter choices,
the solutions can be solved explicitly.  In particular, we obtain
two types of exact solutions.  One is the pp-wave propagating in the
AdS$_3$ background, including the one arising at the critical
points. The other is the BPS string (domain wall) solution, which is
dual to some two-dimensional boundary theory with an ultra-violet
conformal fixed point. We obtain the characteristics of the spectrum
using the standard free-scalar analysis.

    Our results are the first tentative approach to understand a
possible quantum supergravity in three dimensions with inevitable higher
derivatives terms in the supergravity sector, but with the standard
dynamics in the matter sector. The existence of a remaining
well-defined massive graviton at the critical points makes bulk
gravity non-trivial, giving rise to a more interesting model to
study quantum gravity. Although the scalar matter multiplet we considered is a rather simple example of a broad class of matter coupldings without higher derivatives (in contrast with string theory), our results nevertheless reveals some general features of these theories. It might be possible to construct a quantum supergravity with inevitable but finite number of higher-derivative terms in the gravity sector coupled with a matter sector of the standard dynamics.

\section*{Acknolwedgement}

We are grateful to Chris Pope, Ergin Sezgin and Zhaolong Wang for
useful discussions and to Miao Li for reading the draft. Y.P.~is
partially supported by an NCFC grant No.10535060/A050207, an NSFC
grant No.10975172, an NSFC group grant No.10821504 and a
GUCAS-BHP-Billiton scholarship.

\end{document}